# Destruction of Molecular Hydrogen During Cosmological Reionization

Zoltán Haiman[1], Martin J. Rees[2], and Abraham Loeb[1]

## ABSTRACT

We investigate the ability of primordial gas clouds to retain molecular hydrogen ($H_2$) during the initial phase of the reionization epoch. We find that before the Strömgren spheres of the individual ionizing sources overlap, the UV background below the ionization threshold is able to penetrate large clouds and suppress their $H_2$ abundance. The consequent lack of $H_2$ cooling could prevent the collapse and fragmentation of clouds with virial temperatures $T_{\rm vir} \lesssim 10^4$ K (or masses $\lesssim 10^8 M_\odot [(1+z_{\rm vir})/10]^{-3/2}$). This negative feedback on structure-formation arises from the very first ionizing sources, and precedes the feedback due to the photoionization heating.

*Subject headings:* cosmology:theory-galaxies:formation-molecular processes-radiative transfer

## 1. Introduction

Lately, there has been a renewed interest in the reionization of the intergalactic medium (IGM) due to its signature on the cosmic microwave background anisotropies (Tegmark, Silk & Blanchard 1994; Aghanim et al. 1996; Loeb 1996). Observations of QSO spectra imply that the IGM has been highly ionized already by the redshift $z \approx 5$ (Schneider, Schmidt & Gunn 1991). Evidence for the presence of an ionizing background radiation with the required magnitude is provided by the proximity effect (Bajtlik, Duncan & Ostriker 1988; Lu, Wolfe, & Turnshek 1991; Bechtold 1994 and references therein). The ionizing sources could have either been stars or quasars (Miralda-Escudé & Ostriker 1990; Cen & Ostriker 1993; Meiksin& Madau 1993; Shapiro, Giroux & Babul 1994; Fukugita & Kawasaki 1994).

In Cold Dark Matter (CDM) cosmologies the first generation of objects forms at redshifts 10-50. Fragmentation of these objects into stars would result in early reionization. Couchman & Rees (1986) have argued that the heating associated with the photoionization of the IGM would inevitably raise the gas temperature to $T \sim 10^4$K and increase the value of the cosmological Jeans mass well above its initial value of $\sim 10^5 M_\odot$. This effect would tend to suppress further baryonic structure formation until the nonlinear mass scale rises above the new Jeans mass. This negative

---

[1] Astronomy Department, Harvard University, 60 Garden Street, Cambridge, MA 02138, USA

[2] Institute of Astronomy, Cambridge University, Madingley Road, Cambridge CB3 1RZ, England

feedback serves as a buffer that monitors the fraction of baryons converted into massive stars and the metalicity of the IGM as a result of early reionization (Fukugita & Kawasaki 1994), and it could in principle explain the nearly universal metalicity value of $\sim 1\%$ observed in Lyman-$\alpha$ absorption systems (Tytler & Fan 1994; Tytler et al. 1995; Cowie et al. 1995).

The ability of primordial gas clouds with virial temperatures $T_{\rm vir} \approx 10^{2-4}$ K to cool radiatively and fragment after they virialize depends strongly on the existence of molecular hydrogen in them (Haiman, Thoul & Loeb 1996, hereafter HTL96; Tegmark et al. 1996). If the existence of $H_2$ is required in order to produce ionizing sources such as stars or quasars, then it is important to consider the negative feedback that an ionizing background has on the $H_2$ abundance in the universe. As hydrogen molecules can be destroyed by photons below the Lyman limit (Stecher & Williams 1967) to which the universe is transparent before being ionized, it is possible that the feedback due to $H_2$ destruction would precede the feedback due to photoionization heating.

In the early phase of reionization, after the first few sources of ionizing radiation appear, the Strömgren spheres of the individual sources are well separated from each other. During this phase, the ionizing photons from each source are absorbed at the boundary of their local Strömgren spheres, leaving the bulk of the Universe neutral. However, the photons below the Lyman limit escape the Strömgren spheres and are able to dissociate $H_2$ molecules elsewhere. Ionizing sources can therefore affect the $H_2$ abundance at a large distance (Dekel & Rees 1987).

In a previous paper (Haiman, Rees & Loeb 1996; hereafter HRL96) we studied the effect of quasar-like ionizing sources (with a power law spectrum) on the $H_2$ cooling of gas clouds after the universe is reionized. We have found that in very dense clouds ($n \gtrsim 10^3$ cm$^{-3}$) the ionizing radiation enhances the formation of molecules, and results in a net cooling of the gas. Here we examine the effect of the background radiation on the $H_2$ molecules at lower densities, inside clouds that are collapsing when most of the universe is still neutral. In contrast with HRL96, we calculate the spectrum of the background radiation before the reionization epoch. The processed photon spectrum above the Lyman limit is strongly suppressed due to HI continuum absorption throughout the universe. Below the Lyman limit, the spectrum is modulated due to HI line absorption. By estimating the required sizes for clouds to self-shield and retain their $H_2$ molecules, we demonstrate that the $H_2$ molecules inside virialized clouds are easily destroyed before the universe gets reionized. The photodissociation is due to photons softer than the Lyman limit; our results are therefore insensitive to whether the first objects emit a thermal (O-B star) or non-thermal spectrum. The early destruction of $H_2$ may provide a negative feedback that suppresses collapse and fragmentation prior to the feedback due to photoionization heating of neutral hydrogen. These two feedback scenarios could differ appreciably in the redshift at which the IGM became ionized, the predicted number of population III stars at the end of the reionization epoch, and the scales of the aggregates in which they, and hence the first nucleosynthesis products, are formed.

The paper is organized as follows. In § 2 we calculate the characteristic spectrum of the ionizing radiation during the early phase of reionization, when the Strömgren spheres around individual sources do not overlap. In § 3 we use this spectrum to examine the question whether relatively weak fluxes of photons can penetrate large clouds and destroy $H_2$ in them. Finally, § 4 summarizes the cosmological implications of the feedback due to the destruction of $H_2$ molecules.

## 2. Spectrum of UV Background During the Early Phase of Reionization

We consider the first stage of the reionization epoch, when most of the universe is still neutral and the Strömgren spheres around individual sources do not overlap. In order to calculate the spectrum of the UV background outside the Strömgren spheres, we make four assumptions: (i) the ionizing sources turn on simultaneously at the redshift $z_{\rm on}$; (ii) the sources are uniformly distributed in space; (iii) the emitted spectral intensity is a power-law with an index $\alpha$; and (iv) the source population maintains a constant comoving density. Assumption (iii) would be appropriate if the ionizing sources were mainly accreting black holes; our conclusions, however, depend primarily on the emission of photons in the 11.18-13.6 ev range, and therefore apply with little change if the radiation comes from O and B stars.

The spectrum of the background radiation is suppressed above the Lyman limit through absorption by the column of neutral hydrogen and helium between the source and the observer. Photons below the Lyman limit will freely escape the Strömgren spheres around their sources. Since $H_2$ is dissociated by line absorption into the Lyman and Werner lines of molecular hydrogen between 11.18-14.67 eV, special care is needed in calculating the spectrum in this range.

The cumulative flux (in units of erg cm$^{-2}$ s$^{-1}$ ster$^{-1}$ Hz$^{-1}$) observed at the frequency $\nu$ and redshift $z_{\rm obs}$ due to sources with a comoving emission coefficient $j(\nu, z)$ (in units of erg cm$^{-3}$ s$^{-1}$ ster$^{-1}$ Hz$^{-1}$) between the redshifts $z_{\rm obs}$ and $z_{\rm on}$ can be written as

$$J(\nu, z_{\rm obs}) = \frac{c}{H_0} \int_{z_{\rm obs}}^{z_{\rm on}} j(\nu_z, z) \frac{\exp\{-\tau(z)\}dz}{(1+z)^2(1+q_0 z)^{1/2}}, \qquad (1)$$

where $\nu_z \equiv \nu(1+z)/(1+z_{\rm obs})$, and $\tau(z)$ is the optical depth due to neutral hydrogen and helium between the observer's redshift $z_{\rm obs}$ and the redshift $z$,

$$\tau(z) = \frac{c}{H_0} \int_{z_{\rm obs}}^{z} \frac{\kappa(\nu_z, z)dz}{(1+z)^2(1+q_0 z)^{1/2}}, \qquad (2)$$

$$\kappa(\nu, z) = n_{\rm H}(z)\sigma_{\rm H}(\nu) + n_{\rm He}(z)\sigma_{\rm He}(\nu). \qquad (3)$$

Above the hydrogen ionization threshold, equation (1) must be evaluated numerically. Below this threshold, the flux is given by the integral in equation (1), with $\tau = 0$ if we ignore absorption due to lines. Assuming $q_0 = 1/2$ and writing the emission coefficient as $j(\nu, z) = j_0(\nu/\nu_H)^{-\alpha}$, where $\nu_H$ is the hydrogen ionization energy, we obtain in this case

$$J(\nu, z_{\rm obs}) = \frac{cj_0}{H_0} \left(\frac{\nu}{\nu_H}\right)^{-\alpha} \frac{(1+z_{\rm obs})^\alpha}{1.5+\alpha} \left[\frac{1}{(1+z_{\rm obs})^{1.5+\alpha}} - \frac{1}{(1+z_{\rm on})^{1.5+\alpha}}\right]. \qquad (4)$$

In reality, the spectrum below the ionization threshold is processed due to line absorption in the Lyman bands of neutral hydrogen. When the IGM is still neutral, the optical depth across the universe in the resonant Lyman lines is $\sim 10^6$, and so each photon gets absorbed at the redshift when its frequency equals one of the Lyman lines. Hence, photons observed to have a frequency $\nu$ at a redshift $z_{\rm obs}$ could have only been emitted from redshifts that are between the observer and an effective screen located some distance away. The redshift of the screen $z_{\rm screen}$ depends on $\nu_{\rm Lyi}$, the frequency of the nearest Lyman line above $\nu$, through the resonance condition

$$\nu \frac{(1+z_{\rm screen})}{(1+z_{\rm obs})} = \nu_{\rm Lyi}. \qquad (5)$$

We can incorporate the effect of line absorption by replacing the turn-on redshift $z_{\rm on}$ with the screen redshift $z_{\rm screen}$ in equation (4). This is equivalent to multiplying the flux in equation (4) by the modulating factor

$$f = \frac{1 - [(1+z_{\rm obs})/(1+z_{\rm screen})]^{1.5+\alpha}}{1 - [(1+z_{\rm obs})/(1+z_{\rm on})]^{1.5+\alpha}} \qquad (6)$$

So far we have ignored the radiative decay of the excited hydrogen atoms. In reality, each absorbing screen will re-emit line photons. Ly$\alpha$ photons are just re-cycled until they redshift out of resonance, because their destruction rate by the 2s$\rightarrow$1s two-photon decay is much slower than the cosmological expansion [cf. equation 6.114 in Peebles (1993)]. Therefore, there is no net absorption at the redshift corresponding to the Ly$\alpha$ resonance. On the other hand, Ly$\beta$ and higher Lyman line photons are re-emitted as a sum of Ly$\alpha$ and other Balmer or lower line photons. Consequently, the screens corresponding to Ly$\beta$ and higher Lyman lines produce two effects: (i) they block the view of all sources beyond the screen redshift, and (ii) they cascade to the ground state by emitting Ly$\alpha$ and some other, lower-frequency, photons. This results in a sawtooth-modulation of the spectrum in the frequency range from Ly$\beta$ up to the Lyman limit, and in a discontinuous step in the spectrum below the Ly$\alpha$ frequency.

Although it is outside the range of the Lyman and Werner bands of molecular hydrogen, the flux just below the Ly$\alpha$ frequency effects the dissociation rate of $H^-$ and $H_2^+$. To calculate this flux, we assume that each radiative excitation to the 3p state or higher is followed instantaneously by a decay to the ground state with the emission of exactly one Ly$\alpha$ photon (plus other lower energy photons which are not important for our analysis). In addition, we assume that the re-emitted Ly$\alpha$ photons immediately shift out of resonance due to cosmological expansion, and proceed to the observer without further scattering. Under these assumptions, the additional flux at the frequency $\nu$ below Ly$\alpha$ due to Ly$\alpha$ emission from all screens can be written as a sum of terms, each of which being similar to equation (4),

$$\Delta J(\nu, z_{\rm obs}) = \sum_{i=\beta,\gamma,\ldots} \frac{cj_0}{H_0} \left(\frac{\nu}{\nu_H}\right)^{-\alpha} \left(\frac{\nu_{\rm Lyi}}{\nu_{\rm Ly\alpha}}\right)^{1-\alpha} \frac{(1+z_{\rm obs})^\alpha}{1.5+\alpha} \left[\frac{1}{(1+z_\alpha)^{1.5+\alpha}} - \frac{1}{(1+z_i)^{1.5+\alpha}}\right], \quad (7)$$

where $z_\alpha$ is the redshift at which the absorbing screens need to be located in order that their Ly$\alpha$

emission would appear at a frequency $\nu$ at the redshift $z_{\rm obs}$,

$$z_\alpha = \frac{\nu_{\rm Ly\alpha}}{\nu}(1+z_{\rm obs}) - 1, \tag{8}$$

and $z_{\rm i}$ is the highest redshift at which the Lyi photons absorbed at $z_\alpha$ could have been emitted, i.e. the location of a Ly(i+1)-absorbing screen,

$$z_{\rm i} = \frac{\nu_{\rm Ly(i+1)}}{\nu_{\rm Lyi}}(1+z_\alpha) - 1. \tag{9}$$

The factor $(\nu_{\rm Lyi}/\nu_{\rm Ly\alpha})^{1-\alpha}$ in equation (7) accounts for the conversion of a Lyi to a Ly$\alpha$ photon during absorption and re-emission.

Figure 1 shows, as an example, the processed spectrum at $z_{\rm obs} = 25$ assuming the sources turn on at $z_{\rm on} = 35$. The top panel shows that above the Lyman limit the flux is suppressed up to the keV energy range, and the bottom panel presents the sawtooth modulation below the Lyman limit, as well as the discontinuous jump at the Ly$\alpha$ frequency.

The assumption that the UV source distribution is homogeneous remains valid as long as the radius around the observer out to the nearest absorbing screen is larger than the clustering length of the sources. Clearly the assumption breaks down for the highest transitions in the Lyman series. However, our calculation is concerned with the molecular Lyman or Werner lines which are separated from the nearest Lyman line by redshift intervals that are typically $[\Delta z/(1+z)] \gtrsim 10^{-2}$, i.e. much larger than the expected clustering lengths at high redshifts.

One may wonder whether it is possible to detect the reionization spectrum directly. The sawtooth template presented in figure 1 has a characteristic structure that should survive until the present time, provided the breakthrough epoch during which the universe becomes transparent to lines is sudden. Following an abrupt breakthrough epoch (as expected for an unclustered source population), the spectrum below the Lyman limit would be preserved and would show up as a slight modulation on top of the contribution from more recent sources which dominate the observed infrared background. The frequencies at which this sawtooth modulation appears depends on the reionization redshift, and its detection would provide invaluable information about the reionization epoch. Unfortunately, the amplitude of the sawtooth modulation is small. We estimate this amplitude based on the DIRBE observations (see Bond 1995 for a review), and assuming a constant comoving density of ionizing sources that ionize the universe abruptly at $z = 5$ and produce $J_{21} = 0.1$ at $z = 2$. For these conditions the modulation amplitude is smaller than $10^{-6}$, implying that direct detection of the reionization signal is not feasible.

## 3. Destruction of Molecular Hydrogen

We may now use the spectrum illustrated in Fig. 1 to derive the equilibrium $H_2$ fraction inside overdense regions in the universe. In particular, we would like to find out how massive should a gas cloud be in order to form sufficient $H_2$ to cool on its dynamical time-scale.

In calculating the H$_2$ fraction, we consider a homogeneous cloud of primordial H/He ratio with a density $\rho$ and temperature $T$, which is composed of the nine species H, H$^-$, H$^+$, He, He$^+$, He$^{++}$, H$_2$, H$_2^+$, and e$^-$. The chemical composition of the gas within an overdense region is determined by the interaction between these species and the background radiation. For the present purpose, we assume chemical and ionization equilibrium, which is reasonable for the densities and temperatures of interest. The equilibrium H$_2$ fraction is expected to provide an upper limit on the actual H$_2$ fraction and so our constraints on the destruction of H$_2$ are conservative. The assumption of equilibrium is powerful in that it makes our conclusions independent of the details of hydrodynamic and chemical history of the gas.

As emphasized before, the most important element of the chemical network in the present context is the dissociation of H$_2$ by photons with energies below 13.6eV. The significance of this two-step process was first highlighted by Solomon in 1965 (cf. Field et al. 1966) and subsequently studied quantitatively by Stecher & Williams (1967). Here we calculate the dissociation rate for this process by summing the oscillator strengths for the Lyman and Werner bands of H$_2$ from Allison and Dalgarno (1970), and multiplying it with the dissociation fractions (for the second step in the two-step process) from Dalgarno and Stephens (1970), and with the flux at each molecular line[3]. HTL96 have shown that the collapse of a primordial gas cloud beyond virialization depends on its ability to form H$_2$ and cool radiatively. The size of a cloud that is able to form H$_2$ depends, in turn, on its self-shielding in the Lyman and Werner lines. In order to correct the dissociation rate for self-shielding as a function of H$_2$ column density, we used the self-shielding function from de Jong, Dalgarno & Boland (1980). Other details of the chemistry, including the full list of reactions and rate coefficients that we used, are described in HRL96.

For the power-law spectrum in equations (4) and (7), we choose a power-law index $\alpha = 0.7$, and an upper cutoff of $E = 40$keV for the background flux, as expected from quasar spectra and fits to the X-ray background (Fabian & Barcons 1992). We leave the normalization of the flux as a free parameter, and specify it as $J_{21}$, the flux (without absorption) at the Lyman limit in units of $10^{-21}$ erg s$^{-1}$ cm$^{-2}$ Hz$^{-1}$ sr$^{-1}$.

We calculate the equilibrium fraction of H$_2$ within a UV-irradiated homogeneous gas cloud by integrating the chemical rate equations. In order to construct the equilibrium H$_2$ profile as a function of radius, we subdivide the homogeneous cloud into a large number of thin plane-parallel slabs, and in each slab we integrate the rate equations until equilibrium is reached, taking account of the cumulative optical depth due to the H, He, and the H$_2$ lines. We consider a top-hat overdensity of 200 relative to the background universe, and five different combinations of redshift and temperature. As a plausible example, we first set $z=20$ and $T = 10^3$K. We then repeat the calculation for $z=10$ and 50 keeping $T = 10^3$K, and for $T = 10^{2.2}$ and $10^{3.5}$K keeping $z=20$. We selected these temperatures to bracket the range favorable for $H_2$ cooling; $H_2$ is not formed below $10^{2.2}$K, while it is collisionally dissociated above $\sim 10^{3.5}$K (see HRL96). The background universe is assumed to have $\Omega_b = 0.1$ and $H_0 = 50$ km s$^{-1}$ Mpc$^{-1}$. As both the temperature and the density of the gas are fixed, the only free parameter left in this calculation is the normalization of the flux, $J_{21}$.

---

[3]The effect of the sawtooth modulation on the H$_2$ dissociation rate can be fairly large. In the limit $z_{\rm obs}/z_{\rm on} \to 0$, the modulation decreases the total dissociation rate by a factor of 32.5. This is the maximum effect the sawtooth modulation can introduce, and as $z_{\rm obs}$ approaches $z_{\rm on}$, the decrease in the rate is smaller [cf. eq. (6)].

An example for a typical $H_2$ fraction profile in a cloud is shown in figure 2. This figure shows the variation of the equilibrium $H_2$ fraction $f_{H2}$ as a function of neutral hydrogen column density $N_H$ for a hydrogen density $n_H = 1$ cm$^{-3}$, a temperature $T = 1000$K, and a flux $J_{21} = 0.1$. Based on the processes which determine the $H_2$ fraction, three characteristic regimes may be distinguished. Up to hydrogen column densities of $N_H \sim 10^{19}$ cm$^{-2}$, the most important mechanism for the destruction of $H_2$ is direct photoionization by photons with energies above 15.4eV. At larger $N_H$, the destruction rate is dominated by the Solomon process, and $f_{H2}$ drops. At still larger depths, $N_H \sim 10^{23}$ cm$^{-2}$, the Lyman and Werner lines of $H_2$ become self-shielding, and collisional destruction by fast electrons (supplied mainly by the ionization of helium) dominates. The equilibrium state deep inside the cloud where $N_H \sim 10^{24}$ cm$^{-2}$ is fully molecular, but this state is not likely to be reached within a Hubble time (see discussion in HRL96).

Figure 2 also shows the ratio of the cooling time to the dynamical time $t_{\rm cool}/t_{\rm dyn}$ as a function of depth, where

$$t_{\rm cool} = \frac{(3/2)nk_B T}{\Lambda_{\rm tot}} \quad \text{and} \quad t_{\rm dyn} = \frac{1}{\sqrt{6\pi G\rho}}. \tag{10}$$

The total cooling function $\Lambda_{\rm tot}$ includes photo-ionization heating, $H_2$ cooling, and other terms that are summarized in detail in HRL96. The $H_2$ cooling term was taken from Lepp & Shull 1984. Up to a column density $N_H \approx 10^{22}$ cm$^{-2}$, photo-ionization heating dominates $H_2$ cooling, and the total cooling function, as well as $t_{\rm cool}$ are negative, as shown in figure 2. At $N_H \approx 10^{22}$ cm$^{-2}$, the amount of $H_2$ is sufficiently large so that $H_2$ cooling is effective, and at this depth $t_{\rm cool}/t_{\rm dyn}$ becomes positive and quickly drops below unity.

Using profiles similar to figure 2 for each value of $J_{21}$, we have found the depth $r_{H2}$ at which a sufficient $H_2$ abundance is formed so that the $H_2$ cooling time is shorter than the dynamical time. For each value of $J_{21}$, we then define the minimum cooling mass to be $M_{H2} \equiv 4\pi/3\rho r_{H2}^3$. Figure 3a shows the cooling mass as a function of $J_{21}$ for five different combinations of redshift and temperature. The most probable combination $T = 10^3$ K, $z = 20$ is shown as a thick line. In figure 3b we show the same curves, obtained by calculating the minimum cooling mass without the sawtooth modulation below 13.6eV.

Figures 3a and 3b bracket the range of possible $H_2$ photodissociation rates, and imply that in general the cooling masses need to be exceedingly large, even when $J_{21}$ has a relatively low value. In particular, the proximity effect confirms the existence of a background flux of at least $J_{21} \approx 0.1$ that keeps the IGM fully ionized at $z \approx 5$. A similar, or even higher value of $J_{21}$ is expected at the time of breakthrough during the reionization epoch at higher redshifts, since the density of the IGM is then higher. Figures 3a and 3b show that even at fluxes that would be able to ionize the IGM only partially, $J_{21} \lesssim 0.01$, the mass required to shield $H_2$ against the radiation is large, $M_{H2} \gtrsim 10^9 M_\odot$. This conclusion can be inferred from both figures 3a and 3b, implying that it is, in fact, insensitive to the details of the sawtooth-modulation of the spectrum.

## 4. Cosmological Implications

Our main result is that the mass of a virialized primordial cloud has to be exceedingly high in order to self-shield against even a low level of background radiation. Figures 3a and b show the mass required to form enough $H_2$ molecules to cool the cloud on its dynamical time-scale, *versus* the normalization of the flux $J_{21}$. The minimum cooling masses shown in these figures are indeed high, in all cases they are above $\sim 10^9 \, M_\odot$ already for fluxes that are too low to fully ionize the IGM.

It has been argued in previous works (HTL96; Tegmark et al. 1996) that clouds with virial temperatures $100 \, K \lesssim T_{\rm vir} \lesssim 10^4 \, K$ are able to cool by $H_2$ and continue to collapse after they virialize. These papers, however, did not take into account any external UV radiation. Our present results imply that the appearance of the very first UV sources will suppress the $H_2$ abundance in these objects, and prevent their post-virialization collapse.

The cooling masses we obtained depend primarily on the number of photons in the narrow spectral range of 11.18eV-13.6eV. Our conclusions are therefore insensitive to the detailed shape of the spectrum, and should hold for either stellar (blackbody) or quasar (power-law) spectra. We also note that in our calculation of the cooling mass, we assumed a homogeneous spherical cloud. Although centrally condensed clouds are, in general, more effective than homogeneous clouds in self-shielding their molecules, a steepening of the density profile is unlikely to decrease the value of the required cooling mass by orders of magnitude.

We have presented our results for a range of temperatures and redshifts. Our results are relevant to any model where the initial fluctuations have amplitudes decreasing with scale, so that cosmic structures form hierarchically. Such models predict different turn-on redshifts for the first sources of radiation; in PIB models, turn-on may occur at $z > 100$; in CDM in the range $z = 10 - 50$; for mixed dark matter models the first structures may form still more recently. Molecular cooling is more efficient at high densities, and therefore at larger redshifts; however in all these models it determines the scale of the first objects that condense out and contribute the first injection of heat into the universe. The amount of UV generated by each such object is very uncertain; it depends on the efficiency of star formation, and whether the IMF favors massive stars, or even super-massive objects or black holes. However, we have shown that $H_2$ cooling is quenched even in massive objects before the UV background has reached the level needed to photoionize the entire IGM. Therefore the radiation of the first sources exerts an important feedback on the cosmogonic process: the only clouds able to fragment into new stars are those with virial temperature above $\sim 10^4$K, i.e. with masses $\gtrsim 10^8 M_\odot [(1 + z_{\rm vir})/10]^{-3/2}$), so that isothermal contraction is possible due to atomic line cooling. Clouds that virialize at lower temperatures will, in the absence of a sufficient amount of $H_2$, maintain virial equilibrium and not fragment into stars.

We therefore draw the robust conclusion that the IGM remained predominantly neutral until a sufficient number of objects with virial temperatures above $10^4$ K (masses above $\sim 10^8 M_\odot [(1 + z_{\rm vir})/10]^{-3/2}$) had gone non-linear. Most of the O-B stars or accreting black holes that photoionized the IGM had to form in systems at least as large as this. Formation of such systems would have continued unimpeded until ionization was complete; subsequently $J_{21}$ rises sharply to a value of order unity, when the universe becomes, in effect, a single HII region. The only net cooling of a fully photoionized gas comes from bremsstrahlung, which is less effective than the collisionally-excited line emission from a partially ionized gas (Efstathiou, 1992; Weinberg, Hernquist and Katz 1996, Navarro and Steinmetz 1996). The completion of photoionization may therefore signal another pause in the cosmogonic process, associated with the further decline in the efficiency of cooling, and the consequent further increase in the minimum scale that can collapse.

We thank Alex Dalgarno for discussions on the chemistry, and the referee, Alain Blanchard, for useful comments. AL acknowledges support from the NASA ATP grant NAG5-3085.

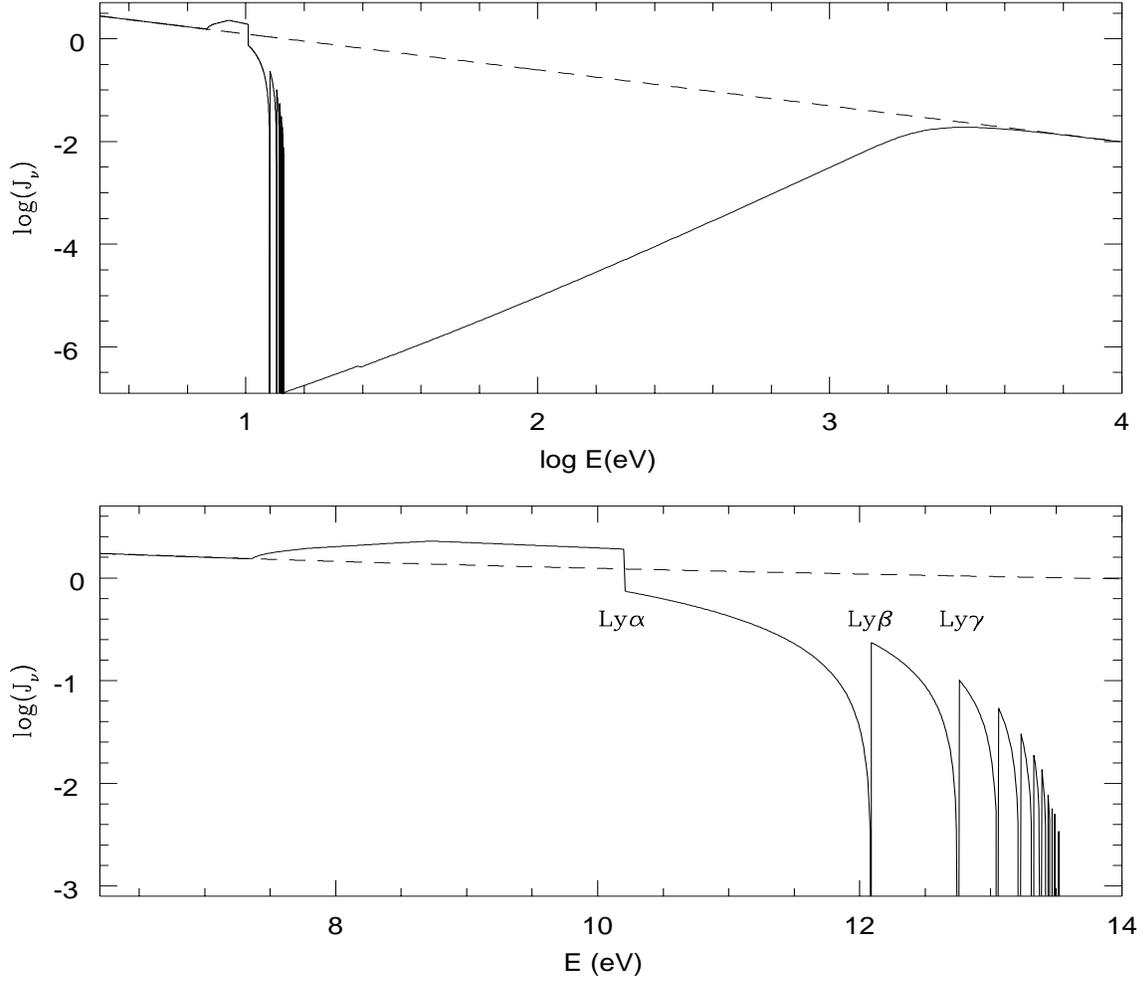

Fig. 1.— The average flux (in units of erg cm$^{-2}$ s$^{-1}$ ster$^{-1}$ Hz$^{-1}$) during the initial phase of the reionization epoch. The upper panel shows that absorption by neutral hydrogen and helium suppresses the flux above 13.6eV up to the keV range. The lower panel shows the sawtooth-modulation due to line absorption below 13.6eV.

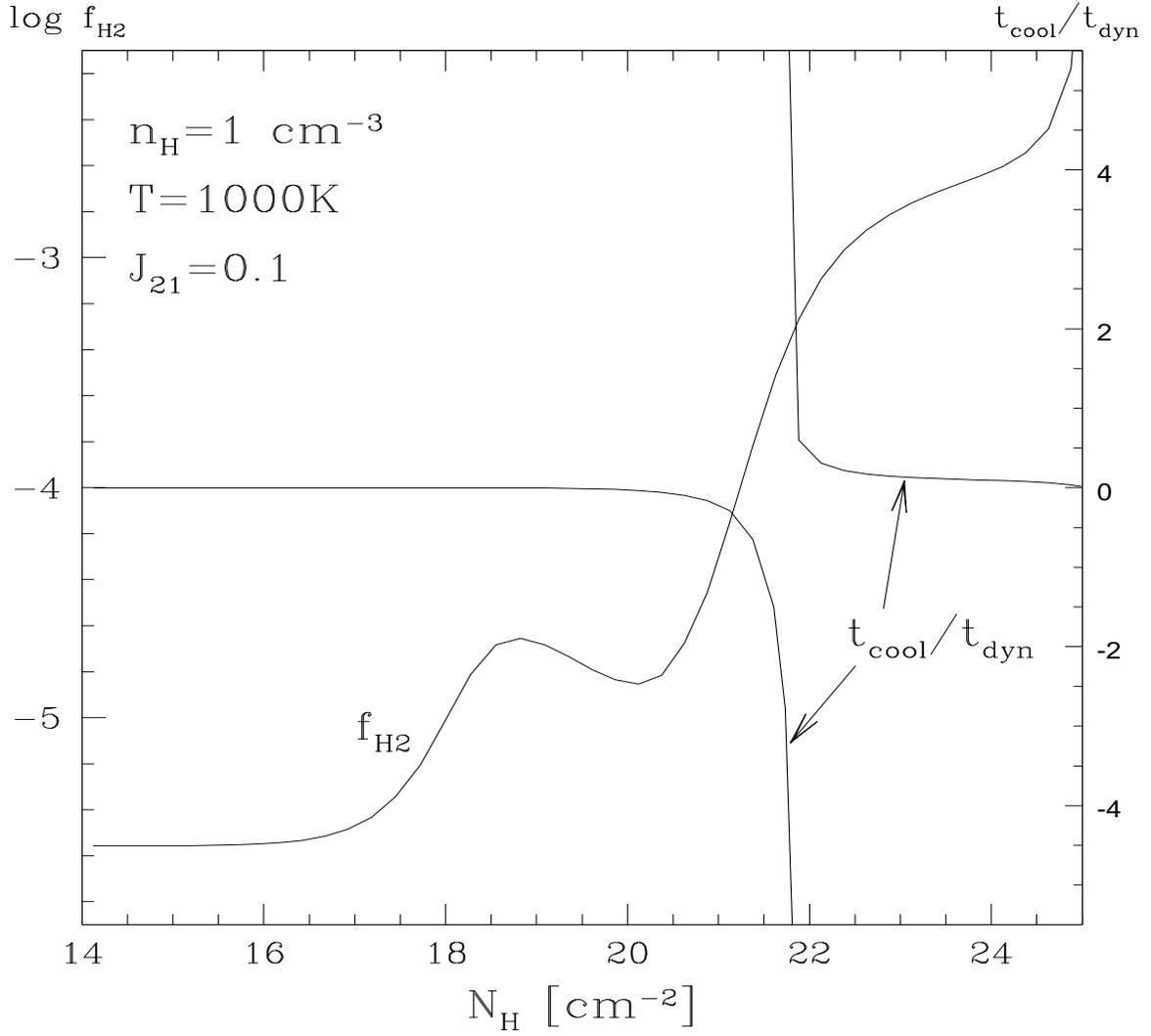

Fig. 2.— The equilibrium profile of the $H_2$ fraction $f_{H_2}$, and the ratio of the cooling time to the dynamical time $t_{cool}/t_{dyn}$. At low column densities, photoionization results in a net heating, this is expressed by $t_{cool} < 0$. At the column density $N_H \sim 10^{21}$ cm$^{-2}$ the $H_2$ Lyman and Werner lines become self-shielding and the molecular fraction starts to rise. When the molecular fraction increases above $10^{-4}$, $H_2$ cooling dominates over photoionization heating, resulting in net cooling ($t_{cool} > 0$). At a slightly larger depth, the $H_2$ cooling time decreases below the dynamical time. The column density at which this occurs, $N_H \sim 10^{22}$ cm$^{-2}$, defines the minimum cooling mass used in § 3.

– 13 –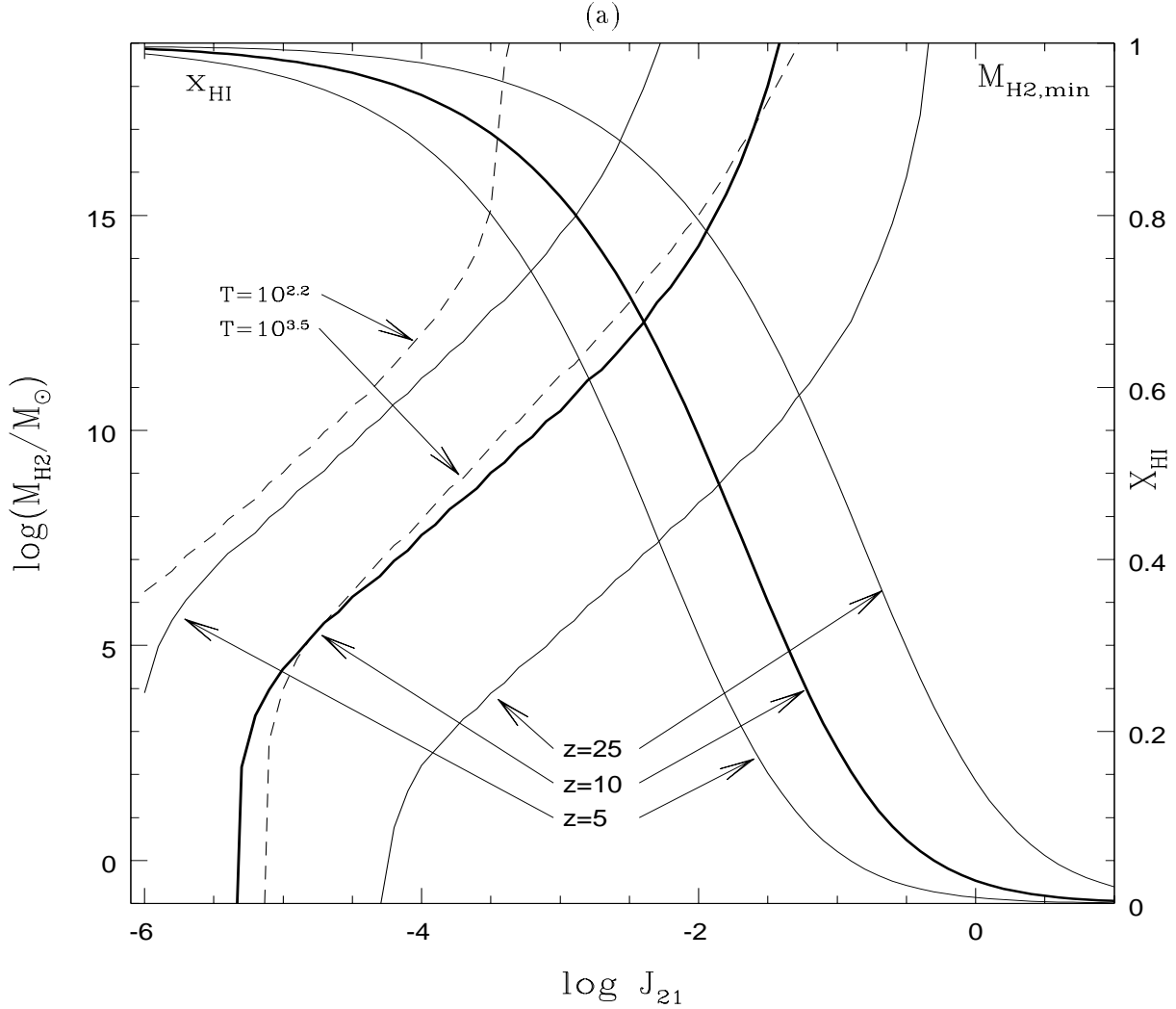

Fig. 3.— (a) The minimum mass $M_{\rm H2}$ of a homogeneous, spherically symmetric cloud which is able to self-shield and form sufficient amount of $H_2$ at its center so that it cools on a dynamical time-scale. The lines show $M_{\rm H2}$ for five different combinations of redshift and temperature. Among these, the intermediate combination $(\log T, z) = (3, 20)$ is shown as a thick line. (b) Same as (a), but with the minimum mass calculated without the sawtooth-modulation of the spectrum below 13.6eV. Note that the cooling masses in (a) and (b) differ only slightly. In both cases, they are high already at fluxes that are too low to ionize the IGM.

– 14 –

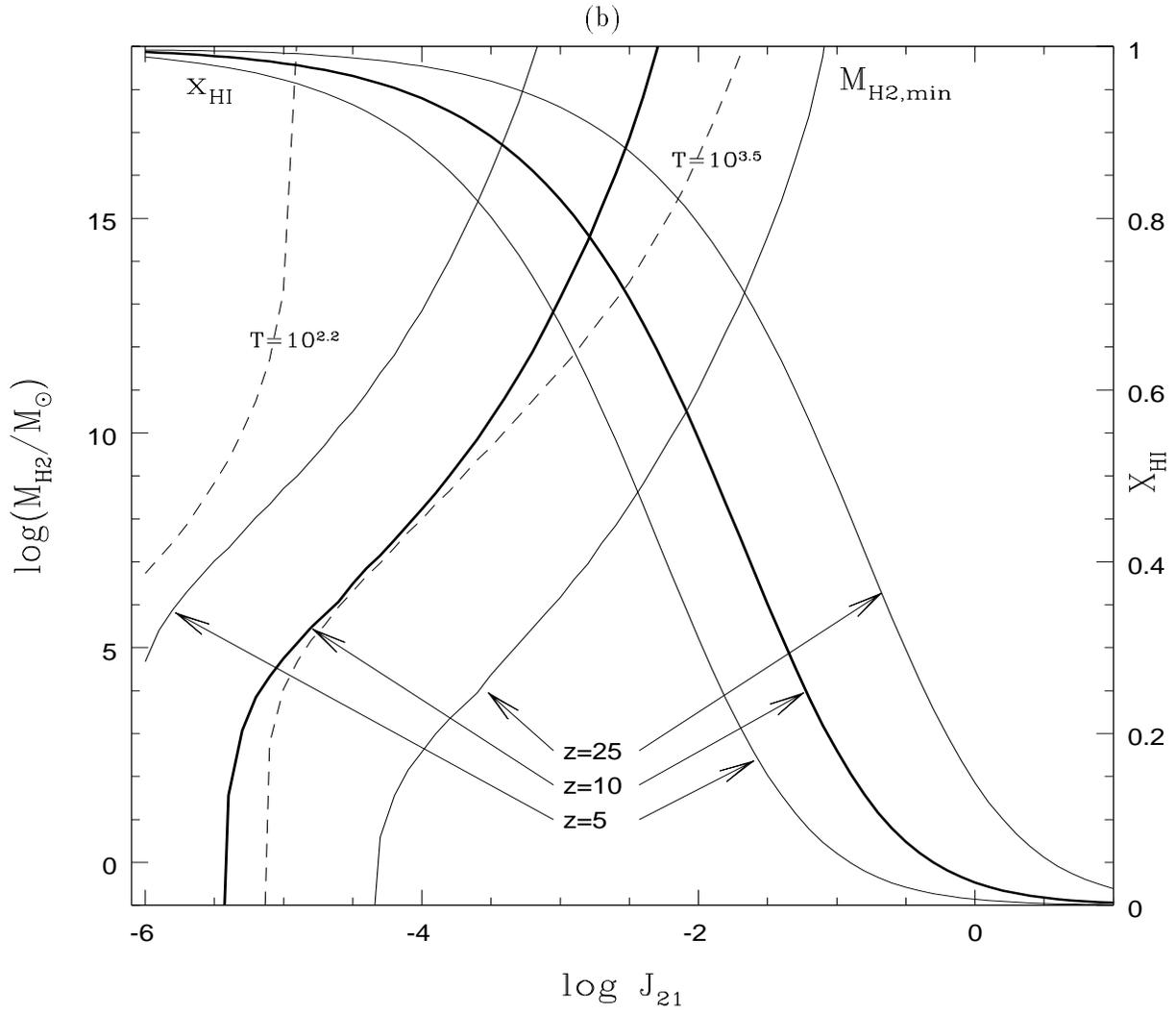